\documentclass[twocolumn,showpacs,superscriptaddress,amsmath,amssymb,aps,pra]{revtex4-1}
\usepackage{bm}
\usepackage{graphicx}
\usepackage{dcolumn}
\usepackage{amsmath,txfonts}
\usepackage{epstopdf}
\usepackage[mathlines]{lineno}
\usepackage{color}
\usepackage[colorlinks=true,linkcolor=blue,citecolor=blue]{hyperref}
\usepackage[normalem]{ulem}

\begin{document}

\title{Third-order exceptional surface in a pseudo-Hermitian superconducting circuit}

\author{Guo-Qiang Zhang}
\email{zhangguoqiang@hznu.edu.cn}
\affiliation{School of Physics, Hangzhou Normal University, Hangzhou 311121, China}

\author{Si-Yan Lin}
\affiliation{School of Physics, Hangzhou Normal University, Hangzhou 311121, China}

\author{Wei Feng}
\affiliation{School of Physics, Hangzhou Normal University, Hangzhou 311121, China}

\author{Yu Wang}
\affiliation{School of Physics, Hangzhou Normal University, Hangzhou 311121, China}

\author{Yang Yu}
\affiliation{School of Physics, Nanjing University, Nanjing 210093, China}

\author{Chui-Ping Yang}
\email{yangcp@hznu.edu.cn}
\affiliation{School of Physics, Hangzhou Normal University, Hangzhou 311121, China}

\begin{abstract}
Compared with an isolated exceptional point, exceptional surfaces in non-Hermitian systems are more robust against environment noises, fabrication errors, and experimental uncertainties. Thanks to this, exceptional surfaces can be applied to enhance the sensitivity of sensors and develop new quantum techniques. Over the past few years, several works have been devoted to studying high-order exceptional surfaces. However, they are restricted to non-Hermitian systems without pseudo-Hermiticity. To date, research on high-order exceptional surfaces in pseudo-Hermitian systems still remains an untouched area. In this work, we propose a pseudo-Hermitian superconducting circuit, which consists of three circularly-coupled superconducting cavities with the balanced gain and loss. We then study the third-order exceptional surface in the proposed circuit. By investigating the eigenvalues, we find that in the parameter space, all third-order exceptional points of the circuit form a third-order exceptional line in the parity-time-symmetric case. When the parity-time-symmetric condition is extended to pseudo-Hermitian conditions, we find more third-order exceptional points, which constitute a third-order exceptional surface in the parameter space. The proposed scheme is universal and can be applied to explore third-order exceptional surfaces in other physical systems, such as optomechanical systems, cavity-magnon systems, and photonic micro-ring systems. This work is of fundamental interest in quantum mechanics and opens a way for studying high-order exceptional surfaces in pseudo-Hermitian systems.

\end{abstract}

\date{\today}

\maketitle

\section{Introduction}
Over the past decade, a number of works have been devoted to studying exceptional points (EPs) in non-Hermitian systems~\cite{Feng17,El-Ganainy18,Ozdemir19,Wiersig20,Bergholtz21,Longhi17,Zhou18,Yang18,Huang22,Jin18,Wiersig14,Chen17,Minganti22,
Zhang21,Song24,Wang20,Minganti21,Wang19,Lu21,Lin25,Zhang22-Zhang,Li23,Arkhipov24}. The EP refers to the spectral singularity of non-Hermitian Hamiltonians, where both $k$ ($k\geq 2$) eigenvalues and eigenstates coalesce~\cite{Heiss12}. The spectral singularity around EPs can give rise to many intriguing phenomena, such as unidirectional invisibility~\cite{Lin11,Peng14,Chatzidimitriou21}, robust wireless power transfer~\cite{Assawaworrarit17,Hao23}, asymmetric mode switching~\cite{Doppler16,Xu16}, phonon lasers~\cite{Jing14,Zhang18,Lu17}, enhancing spontaneous emission~\cite{Lin16}, and coherent perfect absorption~\cite{Sun14,Zhang17,Wang21}. In particular, Quijandr\'{\i}a \emph{et al.} proposed the second-order EP in a superconducting (SC) circuit~\cite{Quijandria18}, and Dogra \emph{et al.} simulated this on the IBM SC quantum processor~\cite{Dogra21}. SC circuits are one of promising platforms for implementing quantum information processing and exploring various exotic phenomena (see Refs.~\cite{Xiang13,Schoelkopf08,Kurizki15} for reviews). In experiments, the second-order EP has been observed in a dissipative SC qubit~\cite{Naghiloo19,Chen21,Chen22,Wang21-1} or a coupled system of two dissipative SC resonators~\cite{Partanen19}. Moreover, Han \emph{et al.} experimentally characterized the exceptional entanglement transition around a second-order EP~\cite{Han23} and the topological invariant for a third-order EP~\cite{Han24} by measuring the dynamical evolutions of SC circuits.

\begin{figure*}
\centering
\includegraphics[width=0.7\textwidth]{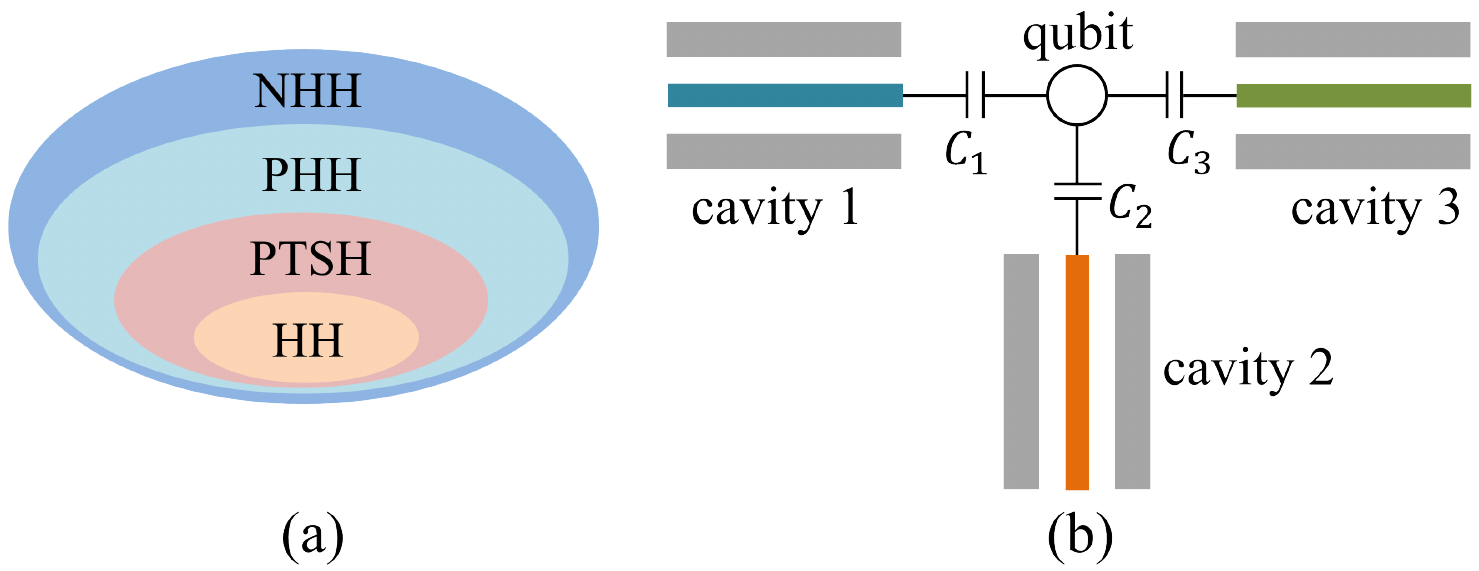}
\caption{(a) The relationship among non-Hermitian, pseudo-Hermitian, $\mathcal{PT}$-symmetric and Hermitian Hamiltonians. In the figure, the non-Hermitian, pseudo-Hermitian, $\mathcal{PT}$-symmetric and Hermitian Hamiltonians are denoted as NHH, PHH, PTSH and HH, respectively. (b) Schematic diagram of the proposed SC circuit, which is composed of three SC coplanar waveguide cavities coupled to a SC qubit (the circle) through capacitances $C_1$, $C_2$, and $C_3$, respectively.}
\label{figure1}
\end{figure*}

In many works, EPs and related applications are associated with the parity-time ($\mathcal{PT}$) symmetry (see, e.g., Refs.~\cite{Liu16,Hodaei17,Feng14,Zhang15,Arkhipov23,Chen22-Zhang,Hu21,Hodaei14}). The $\mathcal{PT}$-symmetric Hamiltonian, satisfying $[H,\mathcal{PT}]=0$, is a special subset of non-Hermitian Hamiltonians~\cite{Bender98}. Usually, the eigenvalues of non-Hermitian Hamiltonians are complex. However, if a non-Hermitian Hamiltonian has the $\mathcal{PT}$ symmetry,
it can also possess an entirely real energy spectrum~\cite{Feng17,El-Ganainy18,Ozdemir19}. It is worth noting that the $\mathcal{PT}$ symmetry can be extended to the $\eta$-pseudo-Hermitian symmetry~\cite{Mostafazadeh02-1,Mostafazadeh02-2,Mostafazadeh02-3}. Here, $\eta$ is a Hermitian invertible operator. Different from the $\mathcal{PT}$-symmetric Hamiltonian, the pseudo-Hermitian Hamiltonian is defined by $\eta H \eta^{-1}=H^\dag$, which has the energy-spectrum properties similar to those of the $\mathcal{PT}$-symmetric Hamiltonian. The relationship among non-Hermitian, pseudo-Hermitian, $\mathcal{PT}$-symmetric and Hermitian Hamiltonians is shown in Fig.~\ref{figure1}(a). In the pseudo-Hermitian case without $\mathcal{PT}$ symmetry, high-order EPs (i.e., $k$th-order EPs with $k \geq 3$) and related applications have been investigated in various physical systems, including cavity-magnon systems~\cite{Zhang19,Zhang23,Hu24}, cavity optomechanical systems~\cite{Xiong21,Xiong22}, multicoil wireless-power-transfer systems~\cite{Hao23}, radio-frequency circuits~\cite{Yin23}, and atom-cavity QED systems~\cite{Li23-Zhong23}.

Recently, research attention has also shifted to the study of exceptional surfaces (ESs)~\cite{Zhong19,Zhou19,Budich19,Grigoryan22,Okugawa19,Zhang19-1,Qin21}. For a $k$th-order ES, every point is a $k$th-order EP. Compared with an isolated EP, ESs have some characteristic properties.
ESs are more robust against environment noises, fabrication errors and experimental uncertainties, which enables ES-based sensors to combine sensitivity and robustness~\cite{Li21,Carlo21,Carlo22,Jiang23}. In optical systems, the ES can grant substantial mastery over system's spectral density of states and band structure, with promising avenues for modulating spontaneous emission rates and amplifying nonlinear effects~\cite{Zhou19}. By engineering ESs via tuning the system parameters, Soleymani \textit{et al.} observed the chiral perfect absorption with quartic lineshape in a waveguide-coupled microresonator system~\cite{Soleymani22}. In addition, the second-order ES has also been studied for controlling spontaneous emission~\cite{Zhong21}, manipulating direction absorption~\cite{Zhong19-1}, topological behaviors~\cite{Tang23,Jia23,Stalhammar21,Wang24}, chaotic dynamics~\cite{Chen22-1}, and optical amplifiers~\cite{Zhong20}.

Similar to the construction of EPs in various $\mathcal{PT}$- and $\eta$-pseudo-Hermitian-symmetric systems, it is critical to combine ESs and $\mathcal{PT}$- and $\eta$-pseudo-Hermitian symmetries. Up to now, the second-order ES has been studied in non-Hermitian systems with different symmetries, e.g.,  $\mathcal{PT}$ symmetry~\cite{Zhou19}, $\eta$-pseudo-Hermitian symmetry~\cite{Grigoryan22} and parity-particle-hole symmetry~\cite{Okugawa19}. Furthermore, high-order ESs (i.e., $k$th-order ESs with $k \geq 3$) were theoretically proposed~\cite{Yang21,Zhang22} and experimentally demonstrated~\cite{Liao23}. Specifically, Ref.~\cite{Liao23} reported a robust on-chip integrated microlaser source based on ESs, where the performances of the microlaser source can be improved by increasing the order $k$ of ESs. However, {\it these works (i.e., Refs.~\cite{Yang21,Zhang22,Liao23}) are restricted to non-Hermitian systems without pseudo-Hermiticity}, where the eigenvalues are complex near ESs. In this context, investigating high-order ESs in pseudo-Hermitian systems is highly desirable, as it may offer novel insights into exploring topological properties and related applications of high-order ESs in non-Hermitian areas~\cite{Zhou19,Grigoryan22,Okugawa19}.

In this work, we propose a pseudo-Hermitian SC circuit and then study the third-order ES in the circuit. As shown in Fig.~\ref{figure1}(b), the proposed circuit consists of three SC coplanar waveguide cavities, which are coupled to a common SC qubit. Due to the large detuning between the qubit and the three SC cavities, we can adiabatically eliminate the degrees of freedom of the qubit and obtain an effective non-Hermitian Hamiltonian, which describes three circularly-coupled SC cavities mediated by a SC qubit [cf.~Sec.~\ref{model-Hamiltonian}]. When the circuit parameters satisfy certain constraints, the circuit possesses the pseudo-Hermiticity, and it has either (i) three real eigenvalues or (ii) one real eigenvalue and two complex-conjugate eigenvalues. In the $\mathcal{PT}$-symmetric case, we find all third-order EPs of the circuit form a line in the parameter space, i.e., third-order exceptional line (EL). When the $\mathcal{PT}$ symmetry is extended to the $\eta$-pseudo-Hermitian symmetry, we find the third-order EL becomes a third-order ES in the parameter space. We further study the energy spectrum of the pseudo-Hermitian SC circuit around third-order EPs. This proposal is universal and can be applied to investigate the third-order ES in other physical systems, such as optomechanical systems~\cite{Xu16}, cavity-magnon systems~\cite{Zhang17}, and photonic micro-ring systems~\cite{Hodaei17}. This work is of fundamental interest in quantum mechanics~\cite{Li2024,Ren2024,Lu2024} and opens a way for studying high-order ESs in pseudo-Hermitian systems~\cite{Zhang19,Zhang23,Hu24,Xiong21,Xiong22,Yin23,Li23-Zhong23}.

This paper is organized as follows. In Sec.~\ref{model}, we introduce the proposed SC circuit and derive the pseudo-Hermitian conditions for the SC circuit. In Sec.~\ref{EL3-PT}, we investigate the third-order EL of the SC circuit in the $\mathcal{PT}$-symmetric case. In Sec.~\ref{ES3-pseudo-Hermitian}, we construct the third-order ES of the SC circuit under the pseudo-Hermitian conditions. We end up with brief discussions and conclusions in Sec.~\ref{discussions}.

\section{Model}\label{model}
As illustrated in Fig.~\ref{figure1}(b), the proposed SC circuit consists of three SC coplanar waveguide cavities, which are indirectly coupled via a SC qubit. Below we first derive the effective Hamiltonian for the three SC cavities by eliminating the degrees of freedom of the qubit in the dispersive regime, and then derive the pseudo-Hermitian conditions for the SC circuit.

\subsection{Effective Hamiltonian of the proposed SC circuit}\label{model-Hamiltonian}

Consider three SC cavities (labeled as cavity 1, cavity 2 and cavity 3, respectively) coupled to a SC qubit [see Fig.~\ref{figure1}(b)]. The total Hamiltonian of the proposed SC circuit contains two parts:
\begin{eqnarray}
H_{\rm tot}=H_{0}+H_{\rm int}.
\end{eqnarray}
Here, $H_{0}$ is the Hamiltonian of bare cavities and qubit given by (hereafter assuming $\hbar=1$)
\begin{eqnarray}\label{H0}
H_{0}=\sum_{n=1}^3 (\omega'_n-i\kappa_n) a_n^\dag a_n + (\omega_{q}-i\gamma)\sigma^+ \sigma^-,
\end{eqnarray}
where $a_n$ and $a_n^\dag$ ($n=1,2,3$) are the annihilation and creation operators of cavity $n$ with angular frequency $\omega'_n$ and loss rate (or gain rate) $\kappa_n$, $\omega_q$ is the transition frequency between ground state $|g\rangle$ and excited state $|e\rangle$ of the qubit, $\gamma$ is the loss rate of the qubit, and $\sigma^- = |g\rangle \langle e|$ and $\sigma^+ = |e\rangle \langle g|$ are the ladder operators of the qubit. In Eq.~(\ref{H0}), $-i\kappa_n a_n^\dag a_n$ and $-i\gamma\sigma^+ \sigma^-$ describe the dissipations of cavity $n$ and qubit, respectively, which have been widely used in studying EPs~\cite{Feng17,El-Ganainy18,Ozdemir19,Wiersig20,Bergholtz21}. Theoretically, this form can be derived using the Langevin equations~\cite{Xiong21}. If cavity $n$ is passive (active), the loss rate (gain rate) is positive (negative), i.e., $\kappa_n > 0$ ($\kappa_n < 0$). Under the rotating-wave approximation, the interaction Hamiltonian between three cavities and qubit is
\begin{eqnarray}
H_{\rm int}=\sum_{n=1}^3 g_{\rm nq} (a_n^\dag \sigma^- + a_n \sigma^+),
\end{eqnarray}
where $g_{\rm nq}$ ($\geq 0$) is the coupling strength between cavity $n$ and qubit. In our work, we assume that the SC circuit is in the strong-coupling regime, where the circuit parameters satisfy
$\{\omega'_n,\,\omega_q\} \gg g_{\rm nq} \gg\{|\kappa_n|,\,\gamma\}$.

When the frequency detuning of cavity $n$ from the qubit is much larger than the coupling strength $g_{\rm nq}$ between them (i.e., $\omega'_n - \omega_q \gg g_{\rm nq}$, corresponding to the dispersive regime), we can use a Fr\"{o}hlich-Nakajima transformation $U= \exp(V)$ on the Hamiltonian $H_{\rm tot}$ to decouple the cavities and the qubit~\cite{Frohlich50,Nakajima55}. The anti-Hermitian operator $V$ needs to satisfy both $V^{\dag}=-V$ and $H_{\rm int} + [H_{0},V ]  \approx 0$. Up to the second order, the transformed Hamiltonian $H_{\rm eff}=U^{\dag}H_{\rm tot}U$ can be approximatively expressed as
\begin{eqnarray}
H_{\rm eff} \approx H_{0}+\frac{1}{2}[H_{I},V].
\end{eqnarray}
If we choose
\begin{eqnarray}
V=-\sum_{n=1}^3 \frac{g_{\rm nq}}{\omega'_n - \omega_q}(a_n^{\dag}\sigma^- - a_n\sigma^+),
\end{eqnarray}
the corresponding effective Hamiltonian is
\begin{eqnarray}\label{efftive-Hamiltonian1}
H_{\rm eff} & = & \sum_{n=1}^3 (\omega'_n-i\kappa_n) a_n^\dag a_n + (\omega_{q}-i\gamma)\sigma^+ \sigma^- \nonumber\\
            &   &  - \sum_{n=1}^3\sum_{m>n}^3 g_{\rm nm}(a_n^\dag a_m + a_n a_m^\dag)\sigma_z \nonumber\\
            &   &  - \sum_{n=1}^3\frac{g_{\rm nq}^2}{\omega'_n - \omega_q}(a_n^\dag a_n \sigma_z+\sigma^+\sigma^-),
\end{eqnarray}
where the effective coupling strength $g_{\rm nm}$ between cavity $n$ and cavity $m$ is
\begin{equation}
g_{\rm nm}=\frac{g_{\rm nq}g_{\rm mq}}{2} \left(\frac{1}{\omega'_n - \omega_q} + \frac{1}{\omega'_m - \omega_q}\right).
\end{equation}
Because the qubit has a large detuning from three SC cavities, we can assume that there is no energy exchange between qubit and cavities. If the SC qubit is initially prepared in its ground state, we can eliminate the degrees of freedom of the qubit
by replacing $\sigma_z$ with $-1$ in Eq.~(\ref{efftive-Hamiltonian1}). Then, the effective Hamiltonian $H_{\rm eff}$ of three SC cavities can be reduced to
\begin{eqnarray}\label{efftive-Hamiltonian2}
H_{\rm eff} = \sum_{n=1}^3 (\omega_n-i\kappa_n) a_n^\dag a_n
                   + \sum_{n=1}^3\sum_{m>n}^3 g_{\rm nm}(a_n^\dag a_m + a_n a_m^\dag),
\end{eqnarray}
with the shifted frequency $\omega_n=\omega'_n + g_{\rm nq}^2/(\omega'_n - \omega_q)$ of cavity $n$.

The effective Hamiltonian in Eq.~(\ref{efftive-Hamiltonian2}) describes three circularly-coupled SC cavities, which are mediated by the SC qubit. Without loss of generality, we assume that cavity 2 is lossy, i.e.,
\begin{eqnarray}\label{}
\kappa_2>0.
\end{eqnarray}
In the absence of cavity 3 (i.e., $g_{13}=g_{23}=0$), the binary system of cavities 1 and 2 has the $\mathcal{PT}$ symmetry when the system parameters satisfy $\omega_1=\omega_2$ and $\kappa_1=-\kappa_2$. This special case has been investigated in Ref.~\cite{Quijandria18}, where the $\mathcal{PT}$-symmetric phase transition at a second-order EP was studied.

\subsection{Pseudo-Hermitian conditions for the SC circuit}
\begin{figure*}
\centering
\includegraphics[width=0.85\textwidth]{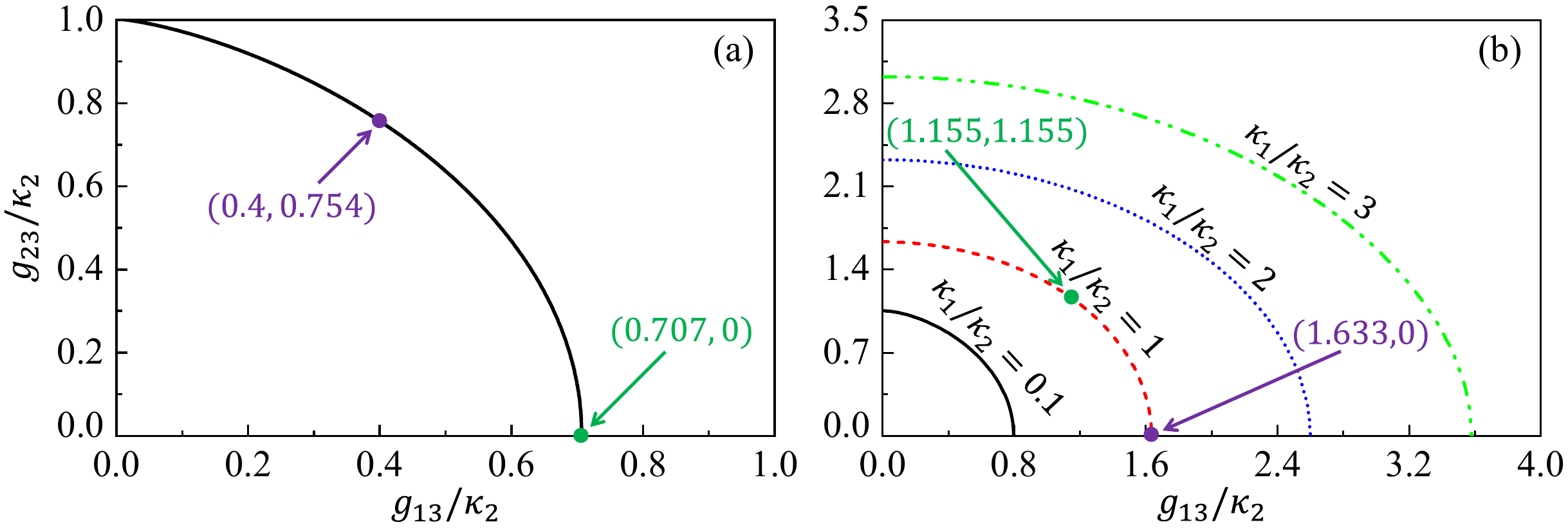}
\caption{(a) Third-order EL in the $\mathcal{PT}$-symmetric case with $\kappa_1=0$, obtained by numerically solving Eq.~(\ref{EL-PT}). (b) Third-order ELs for different $\kappa_1/\kappa_2$, obtained by numerically solving Eq.~(\ref{EP-condition}) under the pseudo-Hermitian conditions in Eq.~(\ref{pseudo-Hermitian-conditions}). Here $\kappa_1/\kappa_2=0.1$ for the (black) solid curve, $\kappa_1/\kappa_2=1$ for the (red) dashed curve, $\kappa_1/\kappa_2=2$ for the (blue) dotted curve, and $\kappa_1/\kappa_2=3$ for the (green) dotted-dashed curve.}
\label{figure2}
\end{figure*}

In this work, we study the third-order ES of the SC circuit under the pseudo-Hermitian conditions. For the convenience of following calculations, we give the corresponding matrix form $H$ of the non-Hermitian Hamiltonian $H_{\rm eff}$ in Eq.~(\ref{efftive-Hamiltonian2}) via the relation $H_{\rm eff}=\alpha^\dag H\alpha$, with $\alpha=(a_1,a_2,a_3)^T$ and
\begin{eqnarray}\label{Hamiltonian-1}
 H=
\left(
  \begin{array}{ccc}
    \omega_1-i\kappa_1 & g_{12} & g_{13} \\
    g_{12} &  \omega_2-i\kappa_2 & g_{23} \\
    g_{13} &  g_{23} & \omega_3-i\kappa_3\\
  \end{array}
\right).
\end{eqnarray}
If the SC circuit possesses the pseudo-Hermiticity, the three eigenvalues of the non-Hermitian matrix $H$ are all real
or one real and other two constituting a complex-conjugate pair~\cite{Mostafazadeh02-1,Mostafazadeh02-2,Mostafazadeh02-3}. This energy-spectrum property of the  pseudo-Hermitian Hamiltonian equals to that the non-Hermitian matrix $H$ and its complex-conjugate matrix $H^*$ have the same eigenvalues~\cite{Zhang19}, i.e., $|H-\lambda I|=|H^*-\lambda I|=0$, where $I$ is an identity matrix and $\lambda$ is the eigenvalue of the non-Hermitian matrix $H$. Using the relation $|H-\lambda I|=|H^*-\lambda I|$, we can obtain the pseudo-Hermitian conditions,
\begin{eqnarray}\label{pseudo-Hermitian-conditions}
\kappa_1+\kappa_2+\kappa_3&=&0,\nonumber\\
\delta_1\kappa_1+\delta_2\kappa_2&=&0,\nonumber\\
g_{12}^2\kappa_3+g_{13}^2\kappa_2+g_{23}^2\kappa_1
          -\delta_1\delta_2\kappa_3+\kappa_1\kappa_2\kappa_3&=&0,
\end{eqnarray}
where $\delta_{1(2)}=\omega_{1(2)}-\omega_3$ is the frequency detuning of cavity 1 (cavity 2) from cavity 3.
Under the pseudo-Hermitian conditions, the characteristic secular equation $|H-\lambda I|=0$ is reduced to
\begin{eqnarray}\label{secular-equation}
a(\lambda-\omega_3)^3+b (\lambda-\omega_3)^2+c (\lambda-\omega_3)+d =0,
\end{eqnarray}
where the four coefficients, $a$, $b$, $c$ and $d$, are given by
\begin{eqnarray}\label{}
a&=&1,\nonumber\\
b&=&-(\delta_1+\delta_2),\nonumber\\
c&=&\delta_1\delta_2-(g_{12}^2+g_{13}^2+g_{23}^2)
        -(\kappa_1\kappa_2+\kappa_1\kappa_3+\kappa_2\kappa_3),\nonumber\\
d&=&\delta_1 g_{23}^2+\delta_2 g_{13}^2-2g_{12}g_{13}g_{23}
          +(\delta_1\kappa_2+\delta_2\kappa_1)\kappa_3.
\end{eqnarray}
Note that Eq.~(\ref{secular-equation}) is a cubic equation for the eigenvalue $\lambda$, which has three roots denoted as $\lambda_{\pm}$ and $\lambda_0$. According to the root discriminant of the cubic equation with one unknown~\cite{Korn68}, the three roots $\lambda_{\pm}$ and $\lambda_0$ are unequally real if the circuit parameters satisfy $\Delta < 0$, where
\begin{eqnarray}\label{}
\Delta \equiv B^{2}-4AC
\end{eqnarray}
is the discriminant, with
\begin{eqnarray}\label{}
A=b^{2}-3ac,~~~
B=bc-9ad,~~~
C=c^{2}-3bd.
\end{eqnarray}
In particular, for $A=B=0$ (satisfying $\Delta=0$) , i.e.,
\begin{eqnarray}\label{EP-condition}
(\kappa_1^2+\kappa_1\kappa_2+\kappa_2^2)(\delta_1^2-3\kappa_2^2)
         +3\kappa_2^2(g_{12}^2+g_{13}^2+g_{23}^2)&=&0,\nonumber\\
(\kappa_1-\kappa_2)\kappa_1 \delta_1^3-18\kappa_2^2g_{13}g_{23}g_{12}
               +\xi\delta_1\kappa_2&=&0,
\end{eqnarray}
$\lambda_{\pm}$ and $\lambda_0$ coalesce to
\begin{eqnarray}\label{EP3-eigenvalue}
\lambda_\pm=\lambda_0=\lambda_{\rm EP3} \equiv \omega_3+\frac{1}{3}(\delta_1+\delta_2),
\end{eqnarray}
which corresponds to the third-order EP of the SC circuit. The coefficient $\xi$ in Eq.~(\ref{EP-condition}) is given by
\begin{eqnarray}\label{}
\xi&=&(g_{12}^2+9\kappa_1\kappa_2)(\kappa_1-\kappa_2)-g_{13}^2(8\kappa_1+\kappa_2)\nonumber\\
   & &+g_{23}^2(\kappa_1+8\kappa_2)+8(\kappa_1^3-\kappa_2^3).
\end{eqnarray}
In the other case of $\Delta=0$ but $A\neq 0$ and $B\neq 0$, only $\lambda_+$ and $\lambda_-$ become coalescent (i.e., $\lambda_+=\lambda_-\neq\lambda_0$), corresponding to the second-order EP of the SC circuit. When $\Delta>0$, only $\lambda_0$ is still real, while $\lambda_+$ and $\lambda_-$ form a complex-conjugate pair.

\section{Third-order exceptional line in the $\mathcal{PT}$-symmetric case}\label{EL3-PT}

In the quantum mechanics, the time-reversal operator $\mathcal{T}$ is usually defined by the complex conjugation operator $\mathcal{K}$, i.e., $\mathcal{T}=\mathcal{K}$~\cite{Bender98}. For our system consisting of three SC cavities, the parity operator $\mathcal{P}$ can be represented as
\begin{eqnarray}\label{}
\mathcal{P}=
\left(
  \begin{array}{ccc}
    1 & 0 & 0 \\
   0 &  0 & 1 \\
    0 &  1 & 0\\
  \end{array}
\right).
\end{eqnarray}
If the proposed SC circuit owns the $\mathcal{PT}$ symmetry, the matrix form $H$ of the non-Hermitian Hamiltonian given in Eq.~(\ref{Hamiltonian-1}) must satisfy
\begin{eqnarray}
[H,\mathcal{PT}]=0,
\end{eqnarray}
which is equivalent to $H=\mathcal{PT}H(\mathcal{PT})^{-1}=\mathcal{P} (\mathcal{K}H\mathcal{K}^{-1})\mathcal{P}^{-1}= \mathcal{P}H^* \mathcal{P}^{-1}$. Using the relation $H=\mathcal{P}H^* \mathcal{P}^{-1}$, we obtain the $\mathcal{PT}$-symmetric conditions given by
\begin{eqnarray}\label{PT-symmetric}
\omega_2=\omega_3,~~~g_{12}=g_{13},~~~\kappa_1=0,~~~\kappa_3=-\kappa_2.
\end{eqnarray}
It can be easily verified that the $\mathcal{PT}$-symmetric conditions are a special case of the pseudo-Hermitian conditions in Eq.~(\ref{pseudo-Hermitian-conditions}). With the $\mathcal{PT}$ symmetry, the conditions of third-order EP given in Eq.~(\ref{EP-condition}) are reduced to
\begin{eqnarray}\label{PT-EP}
\delta_1^2+6g_{13}^2+3g_{23}^2-3\kappa_2^2&=&0,\nonumber\\
9g_{23}g_{13}^2-(4g_{23}^2-g_{13}^2-4\kappa_{2}^2)\delta_1&=&0.
\end{eqnarray}
From the first equation in Eq.~(\ref{PT-EP}), we have $0\leq g_{13}/\kappa_2\leq 1/\sqrt{2}$ and $0\leq g_{23}/\kappa_2\leq 1$. By eliminating $\delta_1$ in Eq.~(\ref{PT-EP}), we find the constraint on the coupling strengths $g_{13}$ and $g_{23}$, given by
\begin{eqnarray}\label{EL-PT}
4g_{23}^2+2g_{13}^2+3\sqrt[3]{4}g_{13}^{4/3}\kappa_2^{2/3}-4\kappa_2^2=0,
\end{eqnarray}
which describes a third-order EL, as plotted in Fig.~\ref{figure2}(a). In the third-order EL, every point corresponds to a third-order EP, where the $\mathcal{PT}$-symmetric phase transition generally occurs~\cite{Feng17, El-Ganainy18,Ozdemir19}. For example, at $g_{13}/\kappa_2=0.707$ and $g_{23}=0$ ($g_{13}/\kappa_2=0.4$ and $g_{23}/\kappa_2=0.754$) indicated by the green dot (purple dot) in Fig.~\ref{figure2}(a), the corresponding three eigenvalues of the circuit coalesce to $\lambda_\pm=\lambda_0=\lambda_{\rm EP3}=\omega_3$ ($\lambda_\pm=\lambda_0=\lambda_{\rm EP3}=\omega_3-0.1923\kappa_2$), cf.~Eq.~(\ref{EP3-eigenvalue}).

Usually, one can observe the EP by measuring the energy spectrum of the system via varying the coupling strength in the experiment, where other system parameters are fixed~\cite{Feng17, El-Ganainy18,Ozdemir19}. In the symmetric case of $\omega_1=\omega_2=\omega_3$ (i.e., the three cavities are on-resonance), the three eigenvalues $\lambda_\pm$ and $\lambda_0$ of the circuit, given by
\begin{eqnarray}\label{pt-eigenvalue}
\lambda_\pm=\omega_3 \pm \sqrt{2g_{13}^2-\kappa_2^2},~~~\lambda_0=\omega_3,
\end{eqnarray}
can be obtained by solving the characteristic equation, i.e., Eq.~(\ref{secular-equation}), under the $\mathcal{PT}$-symmetric conditions given by Eq.~(\ref{PT-symmetric}). According to Eq.~(\ref{pt-eigenvalue}), we find that the SC circuit has three real eigenvalues (one real eigenvalue and two complex-conjugate eigenvalues) in the region $g_{13}/\kappa_2>1/\sqrt{2}$ ($g_{13}/\kappa_2<1/\sqrt{2}$). This indicates that the circuit is in the $\mathcal{PT}$-symmetric phase for $g_{13}/\kappa_2>1/\sqrt{2}$ and the $\mathcal{PT}$-symmetry-breaking phase for $g_{13}/\kappa_2<1/\sqrt{2}$, respectively. Particularly, the three eigenvalues coalesce to $\lambda_\pm=\lambda_0=\lambda_{\rm EP3}=\omega_3$ at $g_{13}/\kappa_2=1/\sqrt{2}$, corresponding to the third-order EP indicated by the green dot in Fig.~\ref{figure2}(a). If we vary the coupling strength $g_{13}$ from $g_{13}/\kappa_2>1/\sqrt{2}$ to $g_{13}/\kappa_2<1/\sqrt{2}$, the circuit will undergo a phase transition from the $\mathcal{PT}$-symmetric phase to the $\mathcal{PT}$-symmetry-breaking phase at a third-order EP with $g_{13}/\kappa_2=1/\sqrt{2}$. In the asymmetric case of $\omega_1\neq \omega_2=\omega_3$ (i.e., the three cavities are off-resonance), the third-order EP still exists, but it is difficult to analytically investigate the eigenvalues of the SC circuit because the expressions of three eigenvalues are cumbersome and not shown here.

\begin{figure}
\centering
\includegraphics[width=0.48\textwidth]{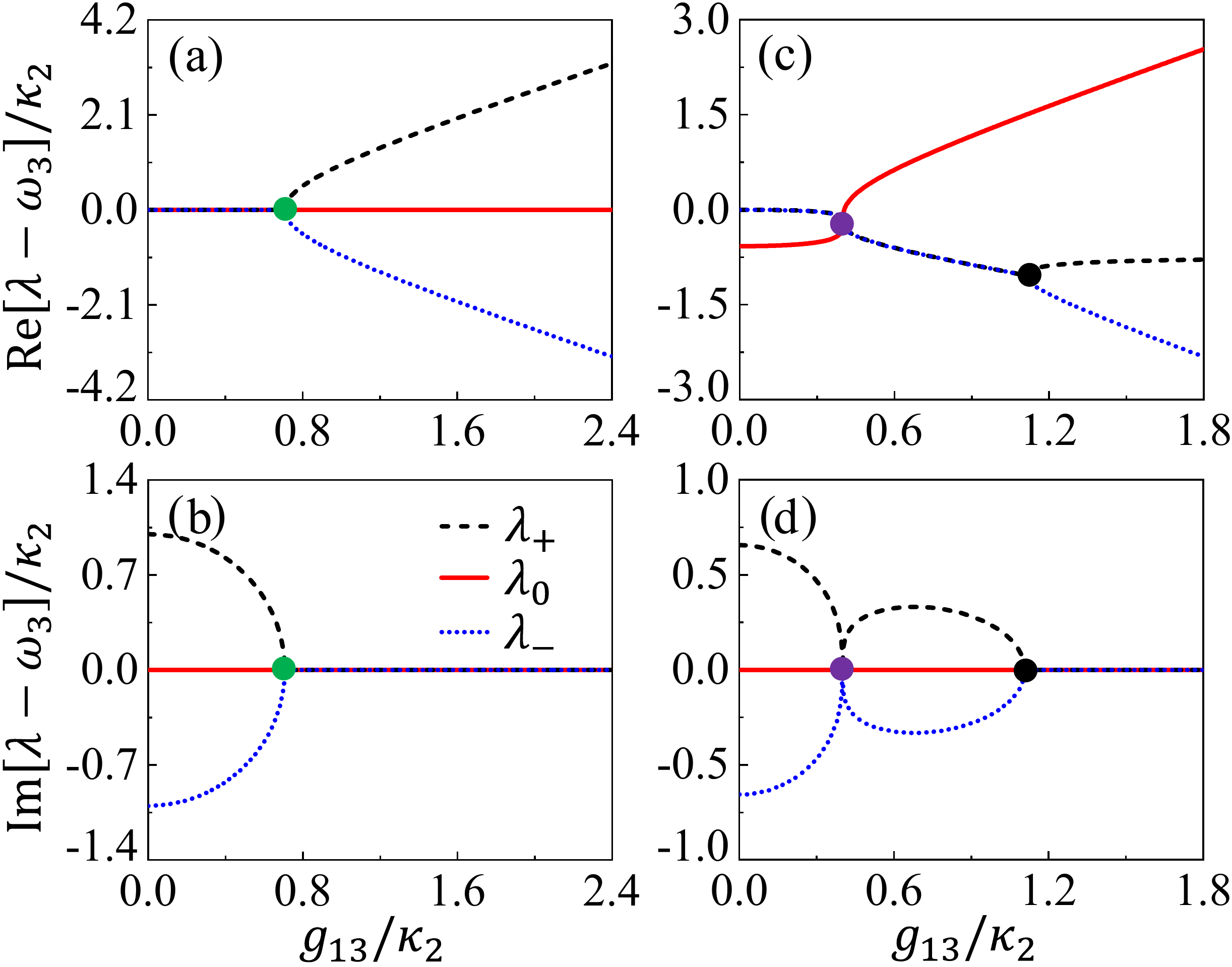}
\caption{The real and imaginary parts of the eigenvalue $(\lambda-\omega_3)/\kappa_2$ as a function of the coupling strength $g_{13}/\kappa_2$ in the $\mathcal{PT}$-symmetric case, where the green dots and purple dots indicate third-order EPs, and the black dots indicate second-order EPs. Here $\delta_1=g_{23}=0$ in (a) and (b), while $\delta_1/\kappa_2=-0.5768$ and $g_{23}/\kappa_2=0.7544$ in (c) and (d). Other parameters are $\kappa_1=0$, $\kappa_3/\kappa_2=-1$, $\delta_2=0$, and $g_{12}=g_{13}$.}
\label{figure3}
\end{figure}

\begin{figure*}
\centering
\includegraphics[width=0.85\textwidth]{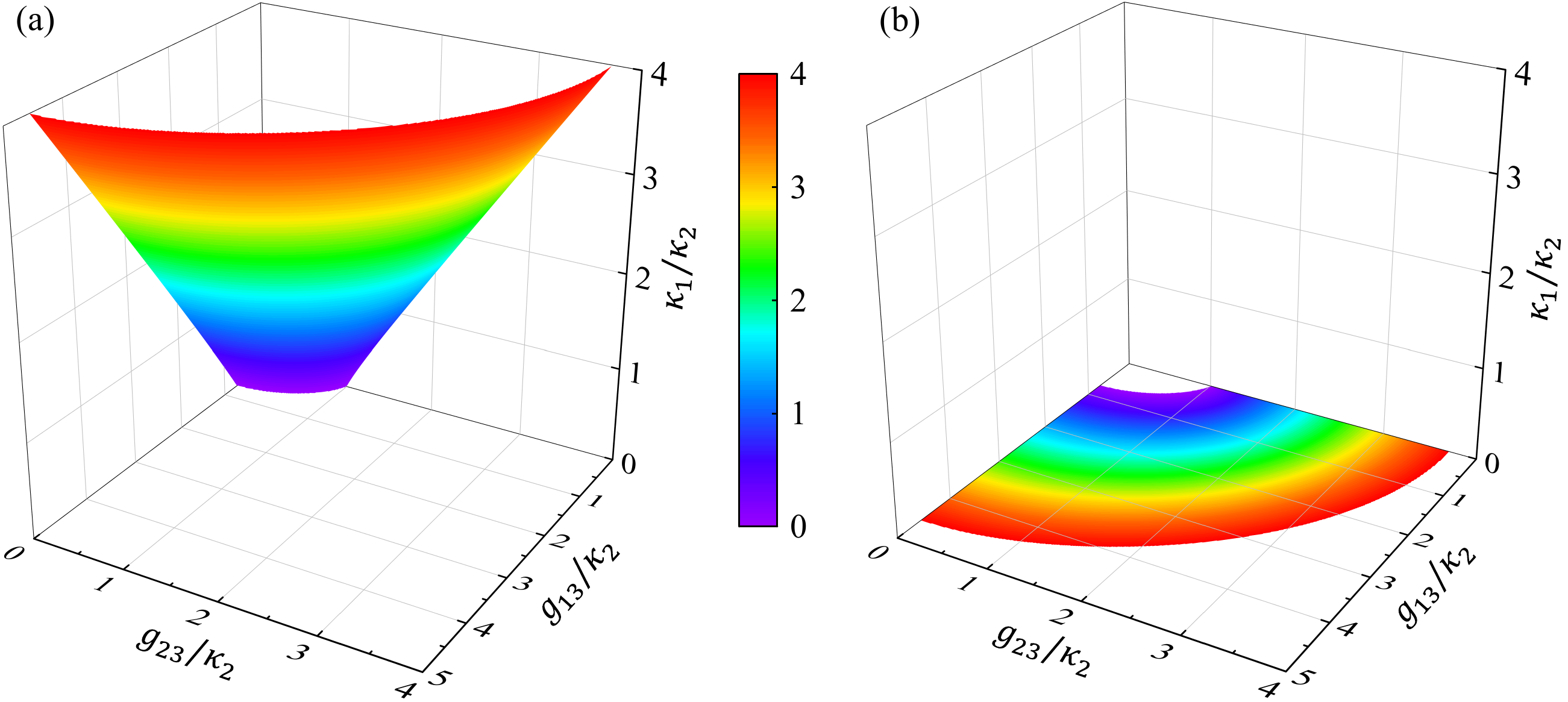}
\caption{(a) Third-order ES obtained by numerically solving Eq.~(\ref{EP-condition}) under the pseudo-Hermitian conditions in Eq.~(\ref{pseudo-Hermitian-conditions}). (b) The projection of the third-order ES depicted in (a) in the space of $g_{13}$ and $g_{23}$. The color in (a) and (b) represents the value of $\kappa_1/\kappa_2$.}
\label{figure4}
\end{figure*}

In Fig.~\ref{figure3}, we display the energy spectrum of the Hamiltonian $H_{\rm eff}$ in Eq.~(\ref{efftive-Hamiltonian2}) versus the coupling strength $g_{13}/\kappa_2$ in the $\mathcal{PT}$-symmetric case. Figures~\ref{figure3}(a) and \ref{figure3}(b) show the real and imaginary parts of the eigenvalues $\lambda_\pm$ and $\lambda_0$ given in Eq.~(\ref{pt-eigenvalue}) versus $g_{13}/\kappa_2$ in the symmetric case (i.e., $\omega_1=\omega_2=\omega_3$). It can be seen that the eigenvalue $\lambda_0$ is real for any values of $g_{13}/\kappa_2$ (see the solid red curves), while other two eigenvalues $\lambda_+$ and $\lambda_-$ are complex (real) when $g_{13}/\kappa_2<0.707$ ($g_{13}/\kappa_2>0.707$) (see the dashed black and dotted blue curves). This means that in the region $g_{13}/\kappa_2>0.707$ ($g_{13}/\kappa_2<0.707$), the circuit is in the $\mathcal{PT}$-symmetric phase ($\mathcal{PT}$-symmetry-breaking phase). In particular, for $g_{13}/\kappa_2=0.707$, the three eigenvalues coalesce together (corresponding to a third-order EP), where the $\mathcal{PT}$-symmetric phase transition occurs. Compared with the symmetric case, the eigenvalues of the SC circuit have significantly different characteristics in the asymmetric case [i.e., $\omega_1\neq \omega_2=\omega_3$; cf.~Figs.~\ref{figure3}(a) and \ref{figure3}(c); Figs.~\ref{figure3}(b) and \ref{figure3}(d)], where the results in Figs.~\ref{figure3}(c) and \ref{figure3}(d) are obtained by numerically solving the characteristic equation [i.e., Eq.~(\ref{secular-equation})] under the $\mathcal{PT}$-symmetric conditions given in Eq.~(\ref{PT-symmetric}). As shown in Figs.~\ref{figure3}(c) and \ref{figure3}(d), the SC circuit can exhibit a $\mathcal{PT}$-symmetric phase transition at the second-order EP [rather than the third-order EP in the symmetric case; cf.~Figs.~\ref{figure3}(a) and \ref{figure3}(b)] with $g_{13}/\kappa_2=1.109$, where only two eigenvalues $\lambda_+$ and $\lambda_-$ (see the dashed black and dotted blue curves) are coalescent. In the $\mathcal{PT}$-symmetry-breaking phase with complex eigenvalues, there exists a third-order EP at $g_{13}/\kappa_2=0.401$, marked by the purple dots in Figs.~\ref{figure3}(c) and \ref{figure3}(d). Note that no $\mathcal{PT}$-symmetric phase transition occurs at this third-order EP.

\section{Third-order exceptional surface under the pseudo-Hermitian conditions}\label{ES3-pseudo-Hermitian}

In the last section, we study the third-order EL in the $\mathcal{PT}$-symmetric case. Here we investigate the third-order ES. When $\kappa_1 > 0$, the SC circuit can own the pseudo-Hermiticity without $\mathcal{PT}$ symmetry. For any given value of $\kappa_1/\kappa_2$ ($> 0$), we can obtain a third-order EL by numerically solving Eq.~(\ref{EP-condition}) under the pseudo-Hermitian conditions given in Eq.~(\ref{pseudo-Hermitian-conditions}). In Fig.~\ref{figure2}(b), we take $\kappa_1/\kappa_2=0.1$, 1, 2, and 3 as an example. At each point in the four third-order ELs, the corresponding three eigenvalues of the SC circuit coalesce to one, such as $\lambda_{\pm}=\lambda_0=\lambda_{\rm EP3}=\omega_3$ at $g_{13}/\kappa_2=g_{23}/\kappa_2=1.155$ (indicated by the green dot) and $\lambda_{\pm}=\lambda_0=\lambda_{\rm EP3}=\omega_3$ at $g_{13}/\kappa_2=1.633$ and $g_{23}=0$ (indicated by the purple dot). If we continuously vary $\kappa_1/\kappa_2$, these third-order ELs will form a surface in the three-dimensional space $(g_{13},g_{23},\kappa_1)$. In Fig.~\ref{figure4}, we plot the distribution of third-order EPs in the space $(g_{13},g_{23},\kappa_1)$ by numerically solving Eqs.~(\ref{EP-condition}) and (\ref{pseudo-Hermitian-conditions}). These third-order EPs form a third-order ES. Note that the third-order ES also contains the third-order EL in the $\mathcal{PT}$-symmetric case with $\kappa_1=0$, cf.~Fig.~\ref{figure2}(a).

\begin{figure}
\centering
\includegraphics[width=0.48\textwidth]{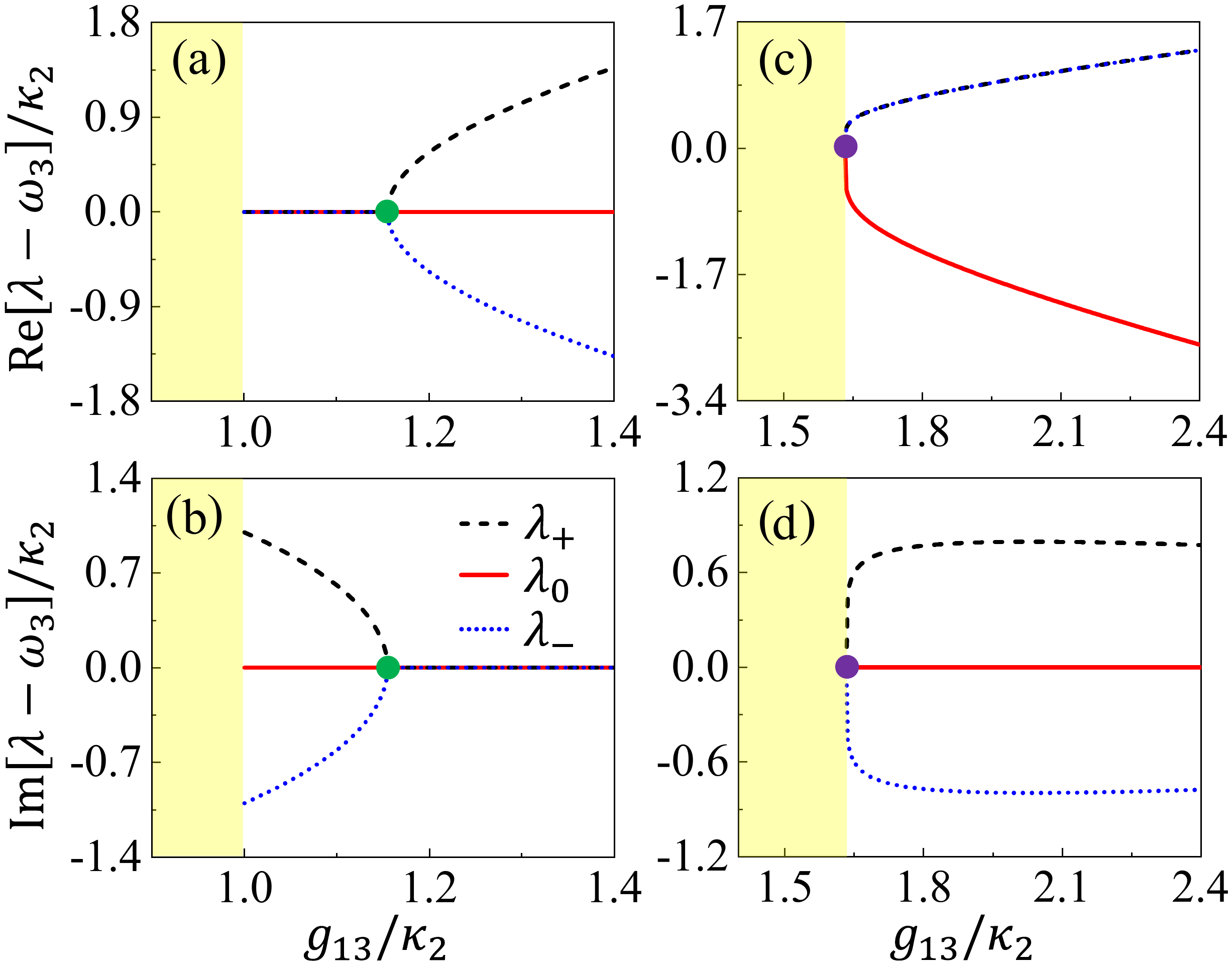}
\caption{The real and imaginary parts of the eigenvalue $(\lambda-\omega_3)/\kappa_2$ versus the coupling strength $g_{13}/\kappa_2$ in the pseudo-Hermitian case without $\mathcal{PT}$ symmetry, where the green dots and purple dots indicate third-order EPs. Note that in the yellow regions, the SC circuit does not have the pseudo-Hermiticity. Here $g_{12}=0$, $g_{23}=g_{13}$ and $\delta_2=\sqrt{g_{13}^2-\kappa_2^2}$ in (a) and (b), while $g_{12}=g_{13}/\sqrt{8}$, $g_{23}=0$ and $\delta_2=\sqrt{3g_{13}^2/8-\kappa_2^2}$ in (c) and (d). Other parameters are $\kappa_1/\kappa_2=1$, $\kappa_3/\kappa_2=-2$ and $\delta_1=-\delta_2$.}
\label{figure5}
\end{figure}

Next, we study the energy spectrum of the SC circuit in the pseudo-Hermitian case with $\kappa_1 > 0$. For the symmetric case of $\kappa_1/\kappa_2=1$, $g_{12}=0$ and $g_{23}=g_{13}$, the pseudo-Hermitian conditions in Eq.~(\ref{pseudo-Hermitian-conditions}) are reduced to
\begin{eqnarray}\label{analytical-pseudo-Hermitian}
\kappa_3=-2\kappa_2,~~~
\delta_1=-\delta_2,~~~
\delta_2=\sqrt{g_{13}^2-\kappa_2^2}.
\end{eqnarray}
The third equation in Eq.~(\ref{analytical-pseudo-Hermitian}) means that there is the pseudo-Hermiticity for the circuit only when the coupling strength $g_{13}$ satisfies $g_{13}^2-\kappa_2^2\geq0$, i.e., $g_{13}/\kappa_2 \geq 1$. This feature is significantly different from the $\mathcal{PT}$-symmetric case, where the SC circuit has the $\mathcal{PT}$ symmetry for any value of $g_{13}$ [cf.~Eq.~(\ref{PT-symmetric}) and related discussions]. Under the pseudo-Hermitian conditions given in Eq.~(\ref{analytical-pseudo-Hermitian}), the three eigenvalues of the SC circuit are given by
\begin{eqnarray}\label{eigenvalue-pseudo}
\lambda_\pm=\omega_3 \pm \sqrt{3g_{13}^2-4\kappa_2^2},~~~\lambda_0=\omega_3.
\end{eqnarray}
In the region $g_{13}/\kappa_2 > 2/\sqrt{3}$, the SC circuit has three real eigenvalues. For $g_{13}/\kappa_2=2/\sqrt{3}$ in particular, the three eigenvalues $\lambda_\pm$ and $\lambda_0$ coalesce to $\lambda_\pm=\lambda_0=\lambda_{\rm EP3}=\omega_3$, corresponding to the third-order EP indicated by the green dot in Fig.~\ref{figure2}(b). While, for $1 \leq g_{13}/\kappa_2<2/\sqrt{3}$, $\lambda_0$ is also real but $\lambda_+$ and $\lambda_-$ become complex. In addition, we take an asymmetric case with $\kappa_1/\kappa_2=1$, $g_{12}=g_{13}/\sqrt{8}$ and $g_{23}=0$. Now the pseudo-Hermitian conditions in Eq.~(\ref{pseudo-Hermitian-conditions}) become
\begin{eqnarray}\label{pseudo-Hermitian-conditions-2}
\kappa_3=-2\kappa_2,~~~
\delta_1=-\delta_2,~~~
\delta_2=\sqrt{3g_{13}^2/8-\kappa_2^2},
\end{eqnarray}
which indicates that the allowed minimal value of the coupling strength $g_{13}$ is $g_{13}= \sqrt{8/3}\kappa_2$ to ensure $3g_{13}^2/8-\kappa_2^2 \geq 0$. In this circumstance, we do not show the cumbersome expressions of three eigenvalues and only give the coalesced eigenvalues $\lambda_\pm=\lambda_0=\lambda_{\rm EP3}=\omega_3$ at $g_{13}/\kappa_2= \sqrt{8/3}$, indicated by the purple dot in Fig.~\ref{figure2}(b).

Furthermore, we plot the energy spectrum of the pseudo-Hermitian SC circuit without $\mathcal{PT}$ symmetry in Fig.~\ref{figure5}. Note that in the yellow regions, no pseudo-Hermiticity exists. Figures~\ref{figure5}(a) and \ref{figure5}(b) display the real and imaginary parts of the eigenvalues $\lambda_\pm$ and $\lambda_0$, given in Eq.~(\ref{eigenvalue-pseudo}), as a function of the coupling strength $g_{13}/\kappa_2$ for the symmetric case of $\kappa_1/\kappa_2=1$, $g_{12}=0$ and $g_{23}=g_{13}$. In the region $1\leq g_{13}/\kappa_2<1.155$, the pseudo-Hermitian circuit has one real eigenvalue (see the solid red curves) and two complex-conjugate eigenvalues (see the dashed black and dotted blue curves). At the critical coupling strength $g_{13}/\kappa_2=1.155$, the three eigenvalues $\lambda_\pm$ and $\lambda_0$ coalesce to $\lambda_\pm=\lambda_0=\lambda_{\rm EP3}=\omega_3$, which is a third-order EP of the pseudo-Hermitian circuit. When $g_{13}/\kappa_2>1.155$, all three eigenvalues are real. In the asymmetric case of $\kappa_1/\kappa_2=1$, $g_{12}=g_{13}/\sqrt{8}$ and $g_{23}=0$, the real and imaginary parts of the eigenvalues $\lambda_\pm$ and $\lambda_0$, obtained by numerically solving Eq.~(\ref{secular-equation}) under the pseudo-Hermitian conditions in Eq.~(\ref{pseudo-Hermitian-conditions-2}), are also shown in Figs.~\ref{figure5}(c) and \ref{figure5}(d), where the eigenvalues exist in the region $g_{13}/\kappa_2 \geq 1.633$. In addition to the third-order EP at $g_{13}/\kappa_2=1.633$ (corresponding to $\lambda_\pm=\lambda_0=\lambda_{\rm EP3}=\omega_3$), the eigenvalue $\lambda_0$ is real (see the solid red curves) while the eigenvalues $\lambda_+$ and $\lambda_-$ are complex-conjugate (see the dashed black and dotted blue curves) for any allowed values of $g_{13}$. Different from Fig.~\ref{figure3}, there are no second-order EPs in Fig.~\ref{figure5}. This discrepancy is primarily attributed to the parameter selection. By carefully choosing appropriate parameters, second-order EPs can also emerge even in the pseudo-Hermitian case without $\mathcal{PT}$ symmetry (see, e.g., Ref.~\cite{Xiong21}). A thorough investigation of this topic is beyond the scope of the current study and not shown here.

\section{Discussions and conclusions}\label{discussions}

If more SC cavities are coupled to a SC qubit, higher-order ESs, such as fourth-order and fifth-order ESs, can also be constructed using a similar methodology outlined in the paper. As the order of ESs increases, the complexity of construction process will grow significantly. By introducing additional symmetries into the pseudo-Hermitian circuit, it is expected to effectively reduce the complexity~\cite{Okugawa19,Grigoryan22}. On the other hand, the Liouvillian EP has garnered extensive attention in research, which is defined via the degeneracy of a Liouvillian superoperator~\cite{Minganti19,Minganti20,Arkhipov20}. This is very different from the Hamiltonian EP, with the degeneracy of a non-Hermitian Hamiltonian. However, existing works exclusively focused on ESs consisting of Hamiltonian EPs~\cite{Zhong19,Zhou19,Budich19,Grigoryan22,Okugawa19,Zhang19-1,Qin21,Li21,Carlo21,Carlo22,Jiang23,
Soleymani22,Zhong21,Zhong19-1,Tang23,Jia23,Stalhammar21,Wang24,Chen22-1,Zhong20,Yang21,Zhang22,Liao23}. In the future, it is a fascinating topic to explore the Liouvillian ES, on which each point is a Liouvillian EP. The Liouville spectrum is relatively complex, which gives rise to the challenge in constructing Liouvillian ESs. One potential approach to addressing this challenge is to incorporate symmetries into the Liouvillian superoperator~\cite{Okugawa19,Grigoryan22}.

Previous studies have primarily focused on high-order ESs without pseudo-Hermiticity~\cite{Yang21,Zhang22,Liao23}. This paper presents the first study of the high-order ES in a pseudo-Hermitian circuit. The pseudo-Hermitian nature of the circuit ensures that all eigenvalues on the third-order ES are real. In the parameter space, certain EPs on the third-order ES mark the boundaries of the phase transition between $\eta$-pseudo-Hermiticity broken and unbroken regions (with complex eigenvalues and real eigenvalues, respectively). In addition, non-Hermitian systems serve as an excellent platform for exploring diverse symmetries. The study of second-order ESs and their topological properties has been actively conducted in non-Hermitian systems with different symmetries, such as $\mathcal{PT}$ symmetry~\cite{Zhou19}, $\eta$-pseudo-Hermitian symmetry~\cite{Grigoryan22} and parity-particle-hole symmetry~\cite{Okugawa19}. Combining high-order ESs with pseudo-Hermiticity holds significant promise for uncovering their topological properties and potential applications. For instance, the third-order ES with pseudo-Hermiticity may further improve the performance of ES-based sensors, as the pseudo-Hermiticity can effectively narrow the spectral linewidths of non-Hermitian systems, and EPs on high-order ESs are more sensitive to weak external perturbations~\cite{Hodaei17,Zhang23}.

Before concluding, we briefly discuss the experimental feasibility of the proposal in SC circuits. In experiments, the typical frequency for a SC cavity can be made as $1-10$~GHz, and the loss rate of a SC cavity is on the order of $0.1-1$~~MHz~\cite{Gu17}. By embedding a SC quantum interference device (SQUID) in a SC cavity, the frequency of the SC cavity can be readily tunable via controlling the bias magnetic flux threading the SQUID loop~\cite{Partanen19}. As for the active SC cavity, its gain rate can be adjusted (ranging from 0 to 6 MHz) via controlling the drive field on the auxiliary SC qubit, which is transversely coupled to the SC cavity~\cite{Quijandria18,Zhang20}. In addition, the coupling strength between SC cavities can be tuned by varying the cavity-qubit coupling strength~\cite{Hoffman11,Whittaker14}. With these accessible technologies, our proposal is experimentally feasible.

In summary, we have studied the third-order ES in the proposed pseudo-Hermitian SC circuit composed of three circularly-coupled SC cavities, where the gain and loss are balanced. Using the energy-spectrum properties of the pseudo-Hermitian Hamiltonian, we have derived the pseudo-Hermitian conditions for the SC circuit. For the $\mathcal{PT}$-symmetric case, we found that all third-order EPs of the circuit are located on a third-order EL in the parameter space. By investigating the eigenvalues of the circuit versus the coupling strength, we have discovered that the circuit exhibits the $\mathcal{PT}$-symmetric phase transition at a third-order EP. Under the pseudo-Hermitian conditions, we found that all third-order EPs form a third-order ES in the parameter space, which contains the third-order EL in the $\mathcal{PT}$-symmetric case. In the pseudo-Hermitian case without $\mathcal{PT}$ symmetry, we have also studied the energy spectrum of the SC circuit around third-order EPs. To the best of our knowledge, this work is the first to study high-order ESs in pseudo-Hermitian systems. These results are of fundamental interest, which may be found applications in enhancing the sensitivity of sensors and developing new quantum techniques.

\section*{Acknowledgments}

This work is supported by the National Natural Science Foundation of China (Grants No.~U21A20436, No.~12205069, No.~12204139 and No.~12404404) and National Key Research and Development Program of China (Grant No.~2024YFA1408900).

\end{document}